\documentclass[12pt]{article}
\usepackage{amsmath, amssymb, graphics}
\usepackage{amsbsy}
\usepackage{amsfonts}
\usepackage{amsmath}
\usepackage{graphicx}
\newcommand{\mathsym}[1]{{}}

\renewcommand{\title}[1]{\vbox{\center\LARGE{#1}}\vspace{5mm}}
\renewcommand{\author}[1]{\vbox{\center#1}\vspace{5mm}}

\def\ci{\cite}

\def \td {\tilde}

\def  \le  {\l_{\rm eff}}

\def \bi{\bibitem}
\def \la {\label}

\def \l {\lambda}
\def\foot{\footnote}

\def \sql {{\sqrt \l}}
\def \adss {$AdS_5 \times S^5$\ }
\def \ads {$AdS_5$\ }

\newcommand{\rf}[1]{(\ref{#1})}
\def \ov {\over}

\def\th{\theta}

\def \ha{{1\ov 2}}

\def \del {\partial}

\def \bi{\bibitem}
\def \la {\label}

\def \l {\lambda}
\def\foot{\footnote}

\def \sql {{\sqrt \l}}

\def \ov {\over}
\def \varpi {{\rm w}}

\def \no {\nonumber }

\def \k {\kappa}
\def \ci {\cite}

\def \ef {\end{document}}

\makeatletter
\renewcommand\section{\@startsection {section}{1}{\z@}%
                                   {-3.5ex \@plus -1ex \@minus -.2ex}%
                                   {2.3ex \@plus.2ex}%
                                   {\normalfont\large\bfseries}}
\renewcommand\subsection{\@startsection{subsection}{2}{\z@}%
                                   {-3.25ex\@plus -1ex \@minus -.2ex}%
                                   {1.5ex \@plus .2ex}%
                                   {\normalfont\normalsize\bfseries}}
\makeatother
\def\b{{\rm b}} 

\def\Tr{{\rm tr}}

\def \L {\Lambda}

\numberwithin{equation}{section} \makeatletter
\@addtoreset{equation}{section}

  \def \a {\alpha}
  \def \b {\beta}

\def \s {\sigma}

\def \sql {\sqrt{\l}}

\newcommand{\be}{\begin{eqnarray}}
\newcommand{\ee}{\end{eqnarray}}

\def \G {W}

\def \diag {{\rm diag}}

\def \const {{\rm const}}

\def \ha {{{\textstyle{1 \ov 2}}}}

\def \bp {\begin{pmatrix}}  \def \ep {\end{pmatrix}}

\def \td {\tilde}

\renewcommand{\theequation}{1.\arabic{equation}}
\def \bi  {\bibitem} 
\def \bea {\be}
\def \eea {\ee}

\def \lcg {light-cone gauge\ } 
\def \lc {light-cone\ } 
\def \ads {AdS\ }
\def \th {\theta}
\def \et {\eta}

\def \ha{{\textstyle{1 \ov 2}}} 
\def \fo{{\textstyle{1 \ov 4}}}

\def \Tr {{\rm Tr}}
\def \rK {{\rm K}}

\def \se {S_{_E}}\def \le {{ L_{_E}}}

\begin{document}

\textwidth 170mm 
\textheight 230mm 
\topmargin -1cm
\oddsidemargin-0.8cm \evensidemargin -0.8cm 
\topskip 9mm 
\headsep9pt

\overfullrule=0pt
\parskip=2pt
\parindent=12pt
\headheight=0in \headsep=0in \topmargin=0in \oddsidemargin=0in

\vspace{ -3cm} \thispagestyle{empty} \vspace{-1cm}
\begin{flushright} 
\end{flushright}
 \vspace{-1cm}
\begin{flushright} Imperial-TP-RR-02-2009
\end{flushright}
\begin{center}

{\Large\bf
Quantum $AdS_5\times  S^5$  superstring
\\ \vskip 0.2cm
 in the  $AdS$ light-cone gauge 
 }
 
 \vspace{0.8cm} {
  S.~Giombi,$^{a,}$\footnote{giombi@physics.harvard.edu} 
  R.~Ricci,$^{b,}$\footnote{r.ricci@imperial.ac.uk } 
  R.~Roiban,$^{c,}$\footnote{radu@phys.psu.edu}
  A.A.~Tseytlin$^{b,}$\footnote{Also at
   Lebedev  Institute, Moscow.\ \ 
   tseytlin@imperial.ac.uk }  
   and  
   C.~Vergu$^{d,}$\footnote{Cristian\_Vergu@brown.edu} 
 }\\
 \vskip  0.5cm

\small
{\em
$^{a}$
Center for the Fundamental Laws of Nature,
Jefferson Physical Laboratory, \\
Harvard University, Cambridge, MA 02138 USA

 $^{b}$  The Blackett Laboratory, Imperial College,
London SW7 2AZ, U.K. 

    $^{c}$ Department of Physics, The Pennsylvania State University,
University Park, PA 16802, USA

$^{d}$ Physics Department, Brown University, Providence, RI 02912, USA
}

\normalsize
\end{center}

 \vskip 0.8cm

 \begin{abstract}
 
We consider the $AdS_5 \times S^5$  superstring 
in the light-cone gauge 
adapted  to a  massless geodesic in $AdS_5$ in the Poincar\'e patch. 
The resulting action has a relatively simple structure which makes it 
a natural  starting point for  various perturbative  quantum computations.
We illustrate the utility of this AdS light-cone gauge  action by computing the 1-loop 
and 2-loop corrections to the null cusp anomalous  dimension
reproducing in a much simpler and efficient way earlier  results obtained in conformal gauge.
This leads to a  further insight into the structure of the
 superstring partition function in non-trivial background. 
\end{abstract}

\newpage

\renewcommand{\theequation}{1.\arabic{equation}}
 \setcounter{equation}{0}
\setcounter{footnote}{0}
\section{Introduction}


The  superstring theory in \adss \ci{mt} has a  complicated form 
and the problem of  finding its exact quantum spectrum
appears to be  a non-trivial one. 
 This is a  Green-Schwarz  \ci{gs} type theory, so 
 a  natural  way to address the question of   its quantization is to use, 
 as in flat space, a   light-cone type gauge. 
 
 There are two  natural choices of  \lcg  in \ads  corresponding to
  the two 
 inequivalent choices of a massless 
 geodesic:  (1) the 
  one running entirely  within $AdS_5$  or   (2)  the one wrapping  a big
  circle of $S^5$.
 The 
 latter choice corresponds
  to expanding near the  ``plane-wave''  vacuum 
 \ci{met,bmn,gkp,ft1,cal,af}  and it  was widely 
   used in  recent studies of the AdS/CFT duality 
 as it is related to  the natural  ``ferromagnetic'' (or ``magnon'')
 spin chain  vacuum 
 on the gauge theory  side (see, e.g., \ci{beis,afr}  for reviews). 
 The resulting  superstring action  has a rather  involved  non-polynomial
  form and thus is not  a simple  starting point for ``first-principles''  quantization.

The 
former choice of light-cone gauge \ci{mt2,mtt}, in which the massless geodesic runs 
entirely within the (Poincar\'e patch of)  $AdS_5$,  leads to a 
simpler action containing terms at most quartic in fermions. 
However,  the importance of the  corresponding light-cone vacuum  state on the 
 gauge-theory side is not immediately clear; for 
 this reason this \lcg choice received previously less  attention
 (see, however,  \ci{at,mttl,aat,kr,ni,hk}).
 In particular, there were practically no  studies of the corresponding quantum theory.

The aim of the present paper is to 
initiate the exploration of the  AdS 
 \lcg  action  at the quantum level. We shall 
 demonstrate   that it leads to 
  a consistent  definition of the quantum  superstring theory 
by repeating the  computations of the 1-loop \ci{ft1,krtt}
 and 2-loop \ci{rt1,rtt} corrections  to the anomaly of the null cusp 
 Wilson line  \ci{kru} or to the leading term in the large spin
 expansion of the energy of the  folded spinning string in AdS space
 in global coordinates.
 We shall  reproduce the previous   results in a much simpler  way, thus
 providing evidence for    the utility  of this \lc gauge.
 \foot{The previous computations for the null cusp were carried out in the 
 conformal gauge where, in contrast to the AdS \lc gauge, 
the bosonic fluctuations mix in a nontrivial way leading to an off-diagonal propagator
and thus complicating the analysis beyond the 1-loop level.
 } 
 We shall verify, in particular,  the cancellation of the  1-loop and 
 2-loop UV divergences. The  cusp anomaly coefficients  we will find 
 match   the strong-coupling Bethe ansatz \ci{bes} 
predictions   \ci{kri,bkk}; this  provides    evidence 
for the quantum integrability of the superstring  formulated in the AdS 
 \lc gauge.\foot{The classical  integrability of the  \adss superstring \ci{bpr}  
on the space of physical degrees of freedom in  the AdS \lcg was discussed  in \ci{aat}.}

 In the 
 companion paper \ci{new} we shall 
 use this approach to evaluate
 the 2-loop correction to  the energy of the 
 folded  spinning  string  with an extra orbital momentum $J$ in $S^5$  \ci{ft1}
 in the scaling limit when $\ln S \gg 1 $ and  ${ J \ov \sql  \ln S}=$fixed). 
 Our \lcg  result, which is 
  different from the one found using  the  conformal gauge \ci{rt2},
 turns out to be in 
 complete agreement with the Asymptotic Bethe Ansatz calculation of \ci{grom} and  with 
 the result of \ci{bas} which generalizes the 
 connection \ci{am} between the  scaling function and the $O(6)$ 
 sigma model.

 A potential  future application of the AdS \lcg approach, which motivates our present interest in it,
 is  the  study of its  near-flat-space expansion  aimed at  
             constructing 
 the inverse string tension expansion of the  energies of   quantum string states 
 with finite  quantum numbers (cf. \ci{rt3}).
 There are several complications along the way.  One of them is the mixing of the 
 ``center-of-mass'' or superparticle \ci{metss,hk}  modes  with the 
  oscillation string modes (they do not  decouple in  curved space).
 Another thorny  issue  is  the realization of the superconformal algebra on excited 
 quantum string states.
While representations will be classified by the same quantum numbers as for
string in AdS  global  coordinates, the use of Poincar\'e coordinates provides new 
possibilities for constructing the excited string modes and realize the superconformal 
algebra.
Indeed, the AdS \lcg  action of \ci{mt2,mtt} is  constructed in the Poincar\'e patch  
and its Hamiltonian $P^-$, whose eigenstates are the quantum string states, is not 
directly related  to the global AdS energy $E$ 
 (moreover, the two operators do not commute).\foot{A
 relation between the  descriptions of the 
 quantum particle states based on $P^-$ and based on $E$ 
  was discussed in \ci{meta,hk}.}
We hope  to  return to these  issues in the future. 
 
 \

The main part of this paper is organized as follows. 
  In section 2 we shall review the 
 \lcg action of  \ci{mt2,mtt}   and discuss its simplest  ``ground state'' solution 
 corresponding to the massless geodesic in $AdS_5$ in Poincar\'e patch.  
 Expanding  near this ground state one finds the same  small fluctuation 
 spectrum -- 8+8 { massless }   bosonic+fermionic   degrees of freedom --
 as in flat space and  the corresponding partition function is trivial. 
 
In section 3 we shall consider   a non-trivial 
solution of the \lcg  action representing an open-string 
euclidean  world surface ending on a null cusp on the boundary 
 of  $AdS_5$ \ci{kru}. 
This is  still a simple  ``homogeneous'' solution -- the coefficients in the 
\lcg action expanded near it  turn out to be  constant. 
As was argued in \ci{krtt}, 
this world surface is  the same (up to an $SO(2,4)$ transformation and a euclidean continuation) 
as the one describing the asymptotic large spin limit \ci{ft1,ftt} 
of the folded  spinning string \ci{gkp}  when the folds approach  the boundary. 
This relation identifies, from a string theory standpoint, the anomaly of a null cusp Wilson line 
and the large spin limit of the anomalous dimension of twist-2 operators,
 the former being given 
by  the string partition function in the  null cusp background.
Expanding the (euclidean analog of the) \lcg action near this null cusp solution 
we find the same small fluctuation spectrum as in \ci{ft1,krtt}
and thus the  same  1-loop  correction to the cusp anomaly function.

In section 4 we shall extend the computation of the string partition function 
in the null cusp background
to 
2-loop order. 
 An important difference compared to the corresponding conformal gauge 
computation \ci{rt1,rtt} is that one of the massive bosonic fluctuations 
acquires a nontrivial (and divergent) 1-point function through a tadpole graph with a 
single fermionic loop. 
 As a result, there are non-vanishing 
 connected but non-1PI contributions to the 2-loop partition function.
Summing them together  with the  1PI contributions leads to cancellation of all UV divergences and 
 reproduces  the Catalan's  constant 
coefficient in the 2-loop  cusp anomaly  found earlier \ci{rt1,rtt}
by a substantially more involved computation  in the conformal gauge.


\renewcommand{\theequation}{2.\arabic{equation}}
 \setcounter{equation}{0}

\section{Superstring action in the AdS \lcg }

Let us begin with a review of the structure of the \adss 
action in the AdS  light-cone gauge   \ci{mt2,mtt}.

We will use the \adss  metric in the Poincar\'e patch ($m=0,1,2,3; \ M= 1,...,6$)
\bea
&&ds^2 = z^{-2} ( dx^m dx_m   + dz^M dz^M)
= z^{-2} ( dx^m dx_m   + dz^2) + du^M du^M \ ,\la{m} \\
&&   x^m x_m =  x^+ x^-  + x^* x   \ , \ \ \ \    \ 
x^\pm = x^3 \pm x^0 \ , \ \ \ \ \ \  x= x^1 + \mathrm{i} x^2 \ , 
\la{x} \\
&& z^M = z u^M \ , \ \ \ \ u^M u^M= 1 \ , \ \ \ \ \ \
z= (z^M z^M)^{1/2} \equiv  e^\phi   \ . \la{z}
\eea
As discussed in \ci{mt2,mtt}, starting with the action of \ci{mt} 
in the above  coordinates 
and fixing the $\k$-symmetry \lcg 
$\Gamma^+ \vartheta^I=0$ on the two 10-d Majorana-Weyl GS spinors $\vartheta^I$,  
one  may also choose the  following  analog of the conformal  gauge 
\be \la{gam}
 \sqrt{ -g}\ g^{\a\b} = {\rm diag} ( - z^2, z^{-2})  \ . 
 \ee
Since  the resulting action   contains  $x^-$ only in the 
$\sqrt{ -g} g^{\a\b} \del_\a x^+ \del_\b x^-$ term it admits a simple solution 
 \be 
 \la{ga} x^+ = p^+ \tau  \ ,   
 \ee
which thus  can be consistently imposed as a constraint additional to \rf{gam} 
to completely fix the two-dimensional diffeomorphism invariance. 
With this choice $x^-$ decouples from the action 
  (it may be determined from  the equations of motion for $g_{\a\b}$
 or the analog of the Virasoro   constraints where it appears only linearly).\foot{
 In general, as in flat space case,  the knowledge of $x^-$ is still required  
 to construct the charges of the symmetry algebra  and vertex operators. 
 Here, however, we will consider  an observable that is determined 
 just by the \lcg action that does not contain $x^-$.}

The resulting AdS \lcg  action  may be  written as 
 \bea
&& S= \ha T \int d \tau \int_0^{2\pi \ell} d \sigma \  L  
\ , \ \ \ \ \ \ \ \ \ \ \ \ \ 
  T= { R^2 \ov 2 \pi  \a'}={ \sql  \ov 2 \pi}  \ , \la{ac}\\
&&
 L =  \dot{x}^* \dot x   + 
\big( \dot z^M  +
 \frac{{\rm i} p^+ }{ z^{2}} z_N \eta_i  (\rho^{MN}){}^i{}_j \eta^j   \big)^2
+  {\rm i} p^+ (\theta^i \dot{\theta}_i +\eta^i\dot{\eta}_i +\theta_i \dot{\theta}^i +\eta_i\dot{\eta}^i )  \nonumber\\[0pt]   
&& \ \ \ \ \ \ \ \ \ \ \
-  \  \frac{(p^+)^2 }{z^{2}}   (\eta^i\eta_i)^2
- \   \frac{1}{z^{4}} ( x'^*x'  + z'^Mz'^M)  \nonumber\\[0pt] 
&&\ \ \ \ \ \ \ \ \ \ \
- \   2 \Big[\ \frac{p^+}{ z^{3}}z^M \eta^i (\rho^M)_{ij}
\big(\theta'^j - \frac{{\rm i}}{z} \eta^j  x'\big)
                    +
\frac{p^+}{ z^{3}}z^M \eta_i (\rho_M^\dagger)^{ij}
\big(\theta'_j + \frac{{\rm i}}{z} \eta_j  x'^*\big)\Big] \ . 
\la{la}
\eea
Here the  $\th^i = (\th_i)^\dagger,   \ \et^i = (\et_i)^\dagger$ \ $(i=1,2,3,4)$ 
transform in the fundamental representation
of $SU(4)$ and parametrize the 
physical fermionic degrees of freedom (the remaining parts of the two 
ten dimensional Majorana-Weyl spinors in the original GS action).\foot{Here $^\dagger$ stands for hermitian  conjugation on the 
Grassmann  algebra, i.e. fermions are complex.}
 The matrices $\rho^{M}_{ij} $ are the off-diagonal blocks of the Dirac 
 matrices in six dimensions in chiral representation   and $\rho^{MN} = \rho^{[M} \rho^\dagger{}^{N]}$
are  the $SO(6)$ generators (see Appendix A).

The action has manifest $SO(6)$ or $ SU(4)$ symmetry.
It is quartic in the $\eta$-fermions and quadratic in the $\theta$-fermions. 
As in the flat space \lcg  GS action,  the factors of $p^+$ can be absorbed by  rescaling 
the  fermions $\th_i$ and $\et_i$
($p^+$  will still  appear in  the expressions for conserved charges).\foot{
The above
 action is related to the one in (1.4),(1.5) in \ci{mtt} by
$\tau \to (p^+)^{-1} \tau$. It is  also related to the action in  (5.29) in \ci{at} by $\sigma \to
T^{-1} \sigma$.}
For generality we introduced the parameter $\ell$ in the range of $\sigma$.
For example, if  we consider  closed string case with  the world-sheet  topology of a cylinder,
before fixing any gauge we can always set $\ell=1$  by a coordinate transformation.\foot{Another   choice is to rescale $\tau$ and $\sigma$ 
to set $x^+=\tau$  and  $\ell=p^+$
as   in the discussion of string interactions in flat space, see also
\ci{mtt}.}
Before gauge fixing the action is also invariant under
$x^m \to k x^m, \  z^M \to k z^M$.  In the gauge-fixed action \rf{la}  this  symmetry becomes
$x \to k x , \  \ \ \  z^M \to k z^M  ,
\ \  \  \s \to  k^{-2} \s  , \ \
\ell \to  k^{-2} \ell ,  $
so we can still set $\ell=1$  by such a  rescaling.
We  can also  consider the open string case 
defined on a strip  or a half-plane  (in the latter case   $\ell=\infty$).

In the next two sections  we shall consider the string path integral with the two-dimensional euclidean version of 
the  action \rf{la},  i.e. with  $e^{\mathrm{i} S}= e^{-\se}$.
The euclidean  action can be formally obtained   from \rf{la} 
by replacing  $\s \to \mathrm{i} \s$ (and assuming that $\ell=\infty$). 
Setting $p^+=1$ 
leads to the action
\bea
&& \se  = \ha T \int d \tau \int^\infty_0 d \sigma \  \le  \ , \la{ae} \\
&&
 \le  =  \dot{x}^* \dot x   
        + \big( \dot z^M  +
        \frac{{\rm i}}{z^{2}} z_N \eta_i  (\rho^{MN}){}^i{}_j \eta^j   \big)^2
        +  {\rm i}  \big(\theta^i \dot{\theta}_i+\eta^i\dot{\eta}_i 
        +\theta_i \dot{\theta}^i +\eta_i\dot{\eta}^i \big)  \nonumber \\[0pt]   
&& \ \ \ \ \ \ \ \ \ \ \
-  \  \frac{1}{ z^{2}}   (\eta^i\eta_i)^2
+ \   \frac{1}{z^{4}} ( x'^*x'  + z'^Mz'^M)  \nonumber\\[0pt] 
&&\ \ \ \ \ \ \ \ \ \ \
+ \   2 {\rm i}  \Big[ \, \frac{1}{z^{3}}z^M \eta^i (\rho^M)_{ij} 
\big(\theta'^j - \frac{{\rm i}}{z} \eta^j  x'\big)
                                     +
                                     \frac{1}{z^{3}}z^M \eta_i (\rho^\dagger_M)^{ij}
\big(\theta'_j + \frac{{\rm i}}{z} \eta_j  x'^*\big)\Big] \ . \la{lae}
\eea
Dropping all $\s$-derivatives in \rf{la} gives 
 the \lc Lagrangian for the \adss superparticle.  
  When quantized \ci{metss,mtt,hk},  it    reproduces
the spectrum of IIB supergravity on $AdS_5 \times S^5$.

The  action \rf{ac},\rf{la} has a natural ``ground state'' --
 the classical solution 
 \bea \la{ge}
 && z = a=\const \ , \\
  && x^+ = p^+ \tau \ , \   \ \ \ \ \  x^- =0  \ ,\ \ \ \ \ \ \ \ \ \
   x^1,x^2,\theta,\eta =0\ .  \la{ss} \ee 
  This  solution -- which is the direct counterpart of the 
  point-like limit of the  superstring in flat space -- 
  describes a massless geodesic parallel to the boundary
  of the Poincar\'e patch running 
  at a distance $a$ from it. 
  It reaches the boundary at spatial infinity
  ($x_3=\infty$).
   The case of $a=\infty$ corresponds to  the massless
  geodesic passing through the horizon  or the center of $AdS_5$ in global coordinates.
  In global coordinates  this   massless geodesic
  is an arc that reaches the boundary of $AdS_5$ (and then reflects back).

  To describe fluctuations near the solution \rf{ge} we may set
  \be  z^M = z^M_0 + \td z^M  \ , \ \ \ \ \ \ \ \ \ 
  z^M_0= e^{\phi_0} u^M_0= a ( 0,0,0,0,0,1)  \ , 
   \ee
  and then the quadratic  fluctuation term in \rf{la} will  take the form 
  (we  rescaled the  fermions  by $p^+$)
 \be
  &&  L_2 =  \dot{x}^* \dot x   +  \dot {\td z}^M \dot {\td z }^M
   - \   a^{-4} ( x'^* x'  + {\td z}'^M {\td z}'^M) \cr
   && \ \
+ \Big[ {\rm i} (\theta^i \dot{\theta}_i +\eta^i\dot{\eta}_i )
- \   2  a^{-2}\eta^i 
(\rho^6)_{ij}\theta'^j + {\rm h.c.}\Big] \ , \la{mas}
\eea
 where
$ \rho^6$ plays the role of  the  charge conjugation  matrix (see \ci{mtt}).
This Lagrangian  describes a collection of 8+8 massless  excitations, i.e. 
it is exactly  the same  action that one  finds from  flat space
GS action  when using a similar parametrization of the 
16 fermionic coordinates;   the only difference is  the presence
of the ``velocity of light''  factor   $c= a^{-2}$.\foot{It can be absorbed
by rescaling  $\s \to  a^{-2} \s$  and then will appear in front of the action
together with string tension 
$T$ as  $T a^{-2}$  and also will rescale the length of the cylinder:
$\ell \to a^2 \ell$.} 

Since the fluctuation spectrum contains 8  massless 2d bosons and 8 massless
 2d fermions  the 1-loop string  partition function or 1-loop
  correction to the 2-d energy vanishes \ci{atr}, in agreement with the fact 
 that the  massless geodesic  should represent   a BPS state.
Including quartic interaction terms in \rf{mas} 
one  may check that the string partition function remains trivial also 
at the 2-loop order. 

Let us comment on the  values of  conserved charges 
on the solution \rf{ge}.
The  expressions for the superconformal 
charges that correspond to the \lcg action \rf{la}  were given in \ci{mt2,mtt} and for  \rf{ge} 
we find that the only non-zero charge densities     are\foot{
We use that here $x^-=0$. The  charge densities are constant so when  integrated they will  have a prefactor
 $ 2 \pi T\ell = \sql\ \ell$  which we omit here.}
\be\la{ki}
P^+=  p^+ \ , \ \ \ \ \   \ \ \ \ \ \  K^+= - \ha a^2  p^+ \ .   \ee
Here   $K^+$  represents a component of  the 
special conformal generator  $K^m$ of $SO(2,4)$;   
its expectation value thus  vanishes 
only if the geodesic  runs directly at the boundary, i.e.  if  $a=0$.

In general, using  the global embedding 
  coordinates  ($\eta^{AB} X_A X_B =
 -X_0^2 + X_1^2 + X_2^2+ X_3^2+ X_4^2 -  X_5^2= -1
$)  a  massless geodesic in $AdS_5$   is  described by (see, e.g., 
 \ci{atr}):
\be X_A = N_A + M_A \tau\ , \ \ \ \ \ \ \ \
\eta^{AB} M_A M_B= \eta^{AB} N _A M_B=0 \ , \ \ \ \
\eta^{AB} N _A N_B=-1   \la{maf}\ee
so that the  $SO(2,4)$ angular momentum   is
$S_{AB} =  N_A M_B - N_B M_A$.
In particular, the choice  when the motion is along the 
third spatial direction and $z=a=1$
corresponds to $N_A= (0,0,0,0,0,1), \ M_A= (p,0,0,p,0,0)$;  
then $S_{50}=S_{53}=  p $.
The relation between $S_{AB}$ generators and standard  basis of conformal group
generators  on $R^{1,3}$ is as follows:
$S_{m4}= \ha (K_m - P_m), \  S_{m5}= \ha (K_m + P_m),\  S_{54} = D,
\ L_{mn} = S_{mn}$ \ ($m,n=0,1,2,3$).
The global AdS energy is $E= S_{05}= \ha (K_0 + P_0)$.
In the present case
then  $K_m=P_m$ ($m=0,3$),  \ $P^0=-P^3=  p$ and thus, up to a trivial  rescaling,  
this is the same
as  in \rf{ki}. 
Thus  the global AdS  energy here  is same as the Poincar\'e
patch  one, $E=P_0=P_3$, but since $K^m$ is non-zero 
this does not represent 
 a conformal primary state.

\def \L {{\cal L}}

\renewcommand{\theequation}{3.\arabic{equation}}
 \setcounter{equation}{0}

\section{Expansion near null cusp  background}

Let us now  turn to another simple but less trivial solution of the 
(euclidean) 
superstring action for which 
 the fluctuation spectrum is
massive  and the full  fluctuation Lagrangian has constant
coefficients -- 
 the null cusp background \ci{kru,krtt}.\foot{As was already mentioned above, the null cusp solution 
  is related \ci{krtt} by an analytic continuation
and a global conformal transformation to the  infinite spin limit
of the  folded string solution \ci{gkp}  which is  indeed a ``homogeneous'' solution \ci{ftt}.}
Starting with \rf{lae}   one finds 
\foot{We thank Martin Kruczenski for  informing us about this  form of the null cusp solution in the AdS light-cone gauge.
}
\be
{z}=\sqrt{\frac{\tau}{\sigma}} \ , \ \ \ \ \  \ \ \ \ \ 
 x_1=x_2=0  \ .  \la{hi}
\ee
In addition we have (we set $p^+=1$)\foot{Here the analog of the Virasoro constraint gives
$\ha (\dot x^+ x'^- +  \dot x^- x'^+)    + \dot z z'=0$.
Let us mention that there  is an even simpler solution --  a (euclidean) 
surface ending on a straight line 
at the boundary,   $x^+= \tau, \ \ x^- =  b^2 \tau, \ \ z= { 1 \ov b \sigma}, \ \ 
 \sigma \in (0, \infty)$. In this case the string partition function should be trivial, 
 i.e. equal to 1 \ci{gdt,mtt}. While simple at first sight, this solution is however not 
 homogeneous -- i.e. the coefficients of the action of small fluctuations are functions 
 of worldsheet coordinates. Thus, carrying out higher order perturbative calculations 
 seems quite involved.
}
\be \la{ji}
x^+ =  \tau \ , \ \ \ \ \ \ \ \ \ \ 
 x^-  = - { 1 \ov 2\sigma} \ , \ \ \ \ \ \ \ \ \ 
 x^+ x^- = - \ha z^2  
\ . \ee
This solution is describing a euclidean  world surface of an open string 
ending on the AdS boundary (we assume that $\tau$ and $\sigma$ 
 change from 0 to $\infty$).  
Since $x^+x^-=0$  at $z=0$ this surface ends on a null cusp.
 
 \def \cW {{\cal W}}
 
Our aim will be to compute the expectation value 
of the corresponding Wilson loop represented \ci{mald, gdt} by the 
euclidean \adss string  path integral with the null cusp boundary 
conditions\foot{Here it is assumed that $x^+$ and $x^-$  are already integrated 
out using  the \lc gauge conditions.
Starting, say, with the Nambu version of superstring 
  action and imposing the orthogonal-gauge conditions on the induced metric \rf{gam} as well as \rf{ga} as
$\delta$-function conditions in the path integral one may then 
get rid of the integrals over $x^+$ and $x^-$.}
\be 
\langle \cW_{\rm cusp} \rangle =  Z_{\rm{string}}
= \int [dx dz d \theta d\eta] \ e^{- \se}  \ . \la{ww} 
\ee
The semiclassical computation of this path integral is based 
on expanding near  the solution \rf{hi}. 
An important feature of this expansion is that it is possible to choose 
the fluctuation fields and worldsheet coordinates 
such that the coefficients of the fluctuation action
become  constant (i.e. independent of $\tau,\sigma$).
Namely, let us define the string coordinate  fluctuations by 
\bea\la{flu}
&& z=\sqrt{\frac{\tau}{\sigma}}\ {\tilde z} \ , \ \ \ \ \ \ \ \ 
{\tilde z} = e^{\tilde \phi}= 1 + \tilde \phi  +\dots~,\ \ \  
 z^M=\sqrt{\frac{\tau}{\sigma}}\ {\tilde z}^M \ , \ \ \ \ 
{\tilde z}^M = e^{\tilde \phi} \tilde u^M  \\
&&
{\tilde u}{}^{a}=  \frac{y^{a}}{1+\frac{1}{4}y^2}~, \ \ \ \ 
{\tilde u}{}^{6} =  \frac{1-\frac{1}{4}y^2}{1+\frac{1}{4}y^2}  \ , \ \ \ \ \ \ \ \ \
~~~~ y^2\equiv \sum_{a=1}^5 (y^a)^2\ , \ \ \ \ \ a=1,...,5 \ , \la{u} \\
&&
x = \sqrt{\frac{\tau}{\sigma}} \ {\tilde x}
~,~~~~~~
\theta=\frac{1}{\sqrt{\sigma}}{\tilde\theta}
~,~~~~~~
\eta=\frac{1}{\sqrt{\sigma}}{\tilde\eta}\ . \la{xx}
\eea
A further redefinition of the worldsheet coordinates $(\tau,\sigma) \to (t,s)$ (we will 
denote by $(p_0,p_1)$ the corresponding two-dimensional momenta, i.e. $(p_0,p_1)=-{\rm i} (\partial_t,\partial_s)$)
\be\la{oo} 
t=\ln \tau~,~~~~~~s=\ln \sigma~,~~~~~~~~~~~dt ds=\frac{d\tau d\sigma}{\tau\sigma}
~,~~~~~~~\tau\partial_\tau=\partial_t
~,~~~~~~~~\sigma\partial_\sigma=\partial_s\ .
\ee
leads then to the following euclidean action \rf{ae}, \rf{lae}:
\be\label{action}
\se &=&\ha T\int dt \int^\infty_{-\infty} ds\ {\cal L}\ ,\\
\label{Lagrangian}
{\cal L}  &=&
\big|\partial_t {\tilde x}+{\ha}{\tilde x}\big|^2+
\frac{1}{{\tilde z}^{4}} \big|\partial_s {\tilde x} -{\ha}{\tilde x}\big|^2 
\\
&+& \big( \partial_t {\tilde z}^M +{\ha} {\tilde z}^M  +
\frac{{\rm i} }{{\tilde z}^2} 
{\tilde \eta}_i  (\rho^{MN}){}^i{}_j {\tilde \eta}^j  {\tilde z}_N \big)^2 
+ \  \frac{1}{{\tilde z}^{4}} \big(\partial_s{\tilde z}^M -{\ha} {\tilde z}^M \big)^2
 \cr
&+&  {\rm i} 
({\tilde \theta}^i \partial_t{\tilde \theta}_i
+{\tilde \eta}^i\partial_t{\tilde \eta}_i + {\tilde \theta}_i
\partial_t{\tilde \theta}^i
+{\tilde \eta}_i\partial_t{\tilde \eta}^i)
-  \ \frac{1}{{\tilde z}^{2}} ({\tilde \eta}^2)^2 
\cr
&+&  2{\rm i}\ \Bigl[\ \frac{1}{{\tilde z}^{3}}{\tilde \eta}^i (\rho^M)_{ij} {\tilde z}^M
(\partial_s{\tilde \theta}^j -{\ha}{\tilde \theta}^j 
- \frac{{\rm i}}{{\tilde z}} {\tilde \eta}^j
 (\partial_s x-{\ha}x))\cr
&&~~~~~~~~~~~~~~~~~~~~~
           +
\frac{1}{{\tilde z}^{3}}{\tilde \eta}_i (\rho^\dagger_M)^{ij} {\tilde z}^M
(\partial_s {\tilde \theta}_j -{\ha} {\tilde \theta}_j
+ \frac{{\rm i}}{{\tilde z}} {\tilde \eta}_j(\partial_sx^* -
{\ha}x^*))\Bigr] \ . 
\ee 
%
Given that  the coefficients in the fluctuation action  are constant, 
we should  find for the partition function in \rf{ww}\foot{The 
presence of extra $\fo$  in the volume factor is due to our choice of unit 
of scale, see below.}
\bea \la{par}
&& Z_{\rm{string}} = e^{-\G} \ , \ \   \ \ \ \ 
\G= \G_0 + \G_1 + \G_2 + ... =  \ha f(\l)  V \ ,\\
 && \ \ \ \ \ \ \ 
V= \fo V_2  , \ \ \  \ \ \ \ \ \  V_2\equiv  \int dt\int  ds \la{voli}\ ,  \eea
where $\G_0=\se$  is the value of the classical action on the solution 
and $ \G_1,\G_2, ...$ are quantum corrections. The cusp 
anomaly function $f(\l)$ has thus the following  inverse string tension expansion 
\be\label{flambda}
f(\lambda)=\frac{\sqrt{\lambda}}{\pi}\Big[1+\frac{a_1}{\sqrt{\lambda}}+\frac{a_2}{(\sqrt{\lambda})^2}+\frac{a_3}{(\sqrt{\lambda})^3}+\cdots\Big]\,.
\ee
To compute the 1-loop coefficient 
 $a_1$   let us consider 
the  quadratic part of the fluctuation Lagrangian which identifies the
spectrum of excitations\foot{Here we used that $\theta^i = \theta^\dagger_i, \
\eta^i = \eta^\dagger_i$ and  ignored a total derivative term.}
\be
{\cal L}_2  &=&(\partial_t {\tilde \phi})^2+(\partial_s {\tilde \phi})^2 +{\tilde \phi}^2 + 
|\partial_t {\tilde x}|^2
     +|\partial_s {\tilde x}|^2
+\ha|{\tilde x}|^2 
+(\partial_t {y}^a)^2+(\partial_s {y}^a)^2\no
 \\
&+&  2{\rm i}\;
({\tilde \theta}^i \partial_t{\tilde \theta}_i
+{\tilde \eta}^i\partial_t{\tilde \eta}_i
)
+2{\rm i}\; {\tilde \eta}^i (\rho^6)_{ij}
        (\partial_s{\tilde \theta}^j -\ha{\tilde \theta}^j)
+
2{\rm i}\; {\tilde \eta}_i (\rho^\dagger_6)^{ij}
(\partial_s {\tilde \theta}_j -\ha {\tilde \theta}_j) ~.
\la{qua}
\ee
We thus find the same mass spectrum as in
conformal gauge \ci{ftt,krtt},  
up to normalization of the mass 
scale.\foot{Here the  classical solution \rf{hi} is 
$z= e^{\frac{1}{2}(t -s)}$ which differs by  a 
rescaling of $s$ and $t$
from the form used in \ci{krtt}. The mass scales in the light-cone and the  
conformal gauges are related as $m^2_{\rm{l.c.}}=\fo\, m^2_{\rm{conf.}}$.}
The bosonic modes are:
one field (${\tilde \phi}$) with $m^2=1$;  two
 fields ($\tilde x,\tilde x^*$) with $m^2=\ha $;  five 
fields ($y^a$) with $m^2= 0$.\foot{
It is interesting to note
 that the analogs of the  first three   modes 
in the  closed string picture (i.e. for fluctuations near the long 
 folded spinning string \ci{gkp} in $AdS_3$ part of $AdS_5$)
 are the angular $AdS_3$  mode ``transverse'' to the profile of the string 
 and the two $AdS_5$ modes ``transverse'' to the $AdS_3$ subspace of the solution \ci{ft1,ftt}.}
As in \rf{mas} (or as in flat space), the   fermions
parametrized by $\theta^i$ and $\eta^i$ 
 have an off-diagonal
kinetic operator but now with non-zero mass terms\foot{
Whenever indices on fermions are not written explicitly we will implicitly assume that
$\theta$ and $\eta$ carry upper indices while $\theta^\dagger$ and $\eta^\dagger$ carry lower indices.}
\be
{\cal L}_{\rm F}&=&{\rm i}\Theta K_F\Theta^T~,  \ \ \ \ \ \ \ 
~~~\Theta=(\theta^i, \theta_i,\eta^i,\eta_i)\equiv(\theta,\theta^\dagger,\eta,\eta^\dagger)
\\[5pt]
K_F&=&
\begin{pmatrix}
0 &  {\rm i} p_0 {\bf 1}_4 & -
({\rm i} p_1+\ha ) \rho^6
& 0 \cr
{\rm i} p_0 {\bf 1}_4 & 0 & 0 & -
({\rm i} p_1+\ha ) \rho^\dagger_6
\cr
+
({\rm i} p_1-\ha ) \rho^6
& 0 & 0 & {\rm i} p_0 {\bf 1}_4  \cr
0 & +
({\rm i} p_1-\ha ) \rho^\dagger_6
& {\rm i} p_0 {\bf 1}_4 & 0
\end{pmatrix}~~.
\label{fermion-K}
\ee
The matrices $\rho^6$ (carrying lower indices) and $\rho_6^\dagger$ (carrying upper indices) 
are related as in Appendix A.
The determinant of the fermionic kinetic operator is $
\det K_F=(p^2+ \fo )^8 $
implying that  all 8 physical fermionic  degrees of freedom
have $m^2=\fo $. The equality of masses of all the fermionic modes is required by
the $SO(6)$ symmetry of the null cusp background \ci{am}.

Having the same mass spectrum   implies the same
(UV finite)  result for the 1-loop partition function as found in 
conformal gauge \ci{ft1,ftt,krtt,rt1}:\foot{Note   that with the 
 choice of normalization we use  here  the
  light-cone and the conformal gauge  volume factors 
are related by  $V_2=4 V_2^{\rm{conf.}}$. 
}
\bea
\G_1= - \ln Z_1 &=&\ha {V_2}\int \frac{d^2p}{(2\pi)^2}\ \Big[\ln(p^2+1)
+2\ln (p^2+\ha)+5\ln p^2-8\ln (p^2+\fo)\Big]\nonumber\\
&=&- {3 \ln 2\ov 8\pi}  V_2 \,,
\la{gaa}
\eea
i.e. we get  $a_1= - 3 \ln 2$  in \rf{flambda}.

In the next section we shall  extend this computation to the 2-loop
 level and show that,  as in the 
conformal gauge \ci{rt1,rtt},  the 2-loop coefficient in \rf{flambda} is 
minus the Catalan's constant 
\be \la{two}
a_2=- {\rm{K}} \ , \ \ \ \ \ \ \ \ \ \ \ \ \ \ \ \
{\rm{K}}=\sum_{k=0}^{\infty}\frac{(-1)^k}{(2k+1)^2}=0.9159...\,.
\ee

\renewcommand{\theequation}{4.\arabic{equation}}
 \setcounter{equation}{0}

\section{Cusp anomaly at 2-loops}

An  important feature of the light-cone gauge action expanded 
near the cusp solution 
is that the bosonic propagator
is diagonal. This is  a  useful simplification 
for higher loop
calculations as we shall now  demonstrate  by the 
explicit computation of the  2-loop coefficient \rf{two} 
in   the cusp anomaly.
Finding  $a_2$ amounts to computing  all
connected  vacuum  Feynman diagrams in the background of the null cusp (\ref{hi}).
We will thus need to expand the light-cone gauge Lagrangian (\ref{Lagrangian}) to the quartic order.\foot{Let us note that  in 
 \cite{rt1} the coefficient $a_2$ was calculated in the conformal gauge by 
 using a T-dual version of the \adss  action in the 
 Poincar\'e patch coordinates. This approach is convenient 
  because the T-dual action is only quadratic in the fermions
\ci{kall}. To get  such an action one must fix the $\kappa$-symmetry by choosing the so called S-gauge \ci{mt2}. 
One may  wonder 
whether it is possible to combine the virtues of the bosonic light cone gauge $x^{+}=\tau$ with the simplicity of the T-dual 
Green-Schwarz action. It turns out  that  the choice of the S-gauge is not compatible with the bosonic light-cone gauge. Indeed,  
the equation of motion for $x^{-}$ would be 
$0=d*dx^{+}+(\delta_{_{IJ}}d*+s_{_{IJ}}d)\bar\vartheta^I \Gamma^+ d\vartheta^J
=(\delta_{_{IJ}}d*+s_{_{IJ}}d)\bar\vartheta^I \Gamma^+ d\vartheta^J$, 
which is in contradiction with the fact that in the S-gauge 
$(\delta_{_{IJ}}d*+s_{_{IJ}}d)\bar\vartheta^I \Gamma^+ d\vartheta^J\ne 0$. As usual, $s_{_{IJ}}=\diag(1,-1)$.
}

\subsection{One-particle irreducible contributions}

\begin{figure}
\begin{center}
\includegraphics[width=80mm]{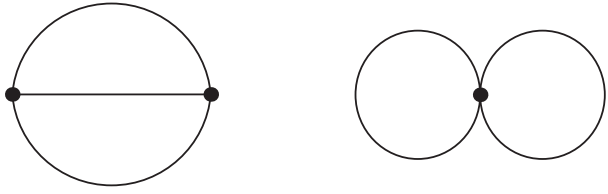}
\parbox{13cm}{\caption{The 2-loop 1PI topologies. The propagators represent either bosonic or fermionic fluctuations.}
\label{1PI}}
\end{center}
\end{figure}

We begin by analyzing the one-particle irreducible contributions to the partition function. At 2-loops they 
correspond to the sunset and double-bubble diagrams, see fig. \ref{1PI}.
The various contributions to the 2-loop part of $\G=-\ln Z_{\rm{string}} $ are obtained from
\be\label{Sint}
\G_2= \langle S_{\rm{int}}\rangle- \ha\langle S_{\rm{int}}^2\rangle_c+
\cdots\ , 
\label{vevSint}
\ee
where $S_{\rm{int}}$ is the interacting part of the action \rf{action},\rf{Lagrangian} containing cubic and  quartic terms. As usual, the Wick contractions are made by inserting the appropriate propagators, and the subscript $c$ indicates that only connected diagrams are to be included. At the  2-loop level, the first term in (\ref{vevSint}) gives the double-bubble diagram 
while the second term gives the sunset diagram as well as the
connected graph with two  tadpoles  which will be discussed in the next subsection.

For the bosonic sunset diagrams we need the following cubic terms from the action,\footnote{
Here we temporarily  set the string tension $T$ to one.
 In the following, we will sometimes  ignore  also the
 obvious  2-d volume factor $V_2$. The dependence on $T$ and $V_2$ is easily  reinserted at the end of the calculation.}
\begin{equation}
\begin{aligned}
&S^{(3)}_{\tilde \phi \tilde x\tilde x^*}=-2\int dt ds\ \tilde\phi\, |\partial_s\tilde x-\ha\tilde x|^2\\
&S^{(3)}_{\tilde \phi^3}= \int dt ds\ \tilde\phi\ [(\partial_t\tilde \phi)^2-(\partial_s\tilde \phi)^2]\\
&S^{(3)}_{\tilde \phi y^2}= \int dt ds\ \tilde\phi\ [(\partial_t y^a)^2-(\partial_s y^a)^2]\,.
\label{bose-3vertices}
\end{aligned}
\end{equation}
The fact that the bosonic propagator is diagonal implies that the sunset graph is simply given by
\be\label{bossunset}
\G_{2\ \rm{bos. sunset}}=-\ha\langle S^{(3)}_{\tilde \phi \tilde x\tilde x}S^{(3)}_{\tilde \phi \tilde x\tilde x}+ S^{(3)}_{\tilde{\phi}^3 }S^{(3)}_{\tilde{\phi}^3}+ S^{(3)}_{\tilde \phi y^2}S^{(3)}_{\tilde \phi y^2}\rangle_{\mbox{\tiny 1PI}}\,.
\ee
All the terms in the above expression can be readily computed. For instance, the 
first term
yields,  in momentum space, 
\be
 &&-2\int d^2p\, d^2q\, d^2r\,\delta^{(2)}(p+q+r) G_{\tilde\phi\tilde\phi}(p)
 \left(q_1^2+\fo\right)G_{\tilde x\tilde x}(q)\left(r_1^2+\fo\right)G_{\tilde x\tilde x}(r)\no  \\
 &&\ =\ -\ha \int d^2p\, d^2q\, d^2r\,\delta^{(2)}(p+q+r) 
 \frac{(1+4q_1^2)(1+4r_1^2)}{(p^2+1)(q^2+\ha)(r^2+\ha)}\,, \la{hh}
\ee
where we have used the propagators for the bosonic fluctuations in \rf{qua}
\be
G_{\tilde x\tilde x^*}(p)=\frac{2}{p^2+\ha}\,,\qquad G_{\tilde\phi\tilde\phi}(p)&=&\frac{1}{p^2+1}\,, \qquad
G_{ y^a y^b}(p)=\frac{\delta^{a b}}{p^2}\,.
\ee
To evaluate the momentum integrals, we employ the same regularization scheme used in \ci{rt1,rtt}. Manipulation 
of tensor structures in the numerators are performed in $d=2$, and the resulting scalar integrals are computed 
in an analytic (e.g., dimensional) regularization scheme in which power divergent contributions are set to 
zero.\foot{The direct cancellation of these divergences amounts to carefully accounting for the contribution of the 
path integral measure and was shown to occur in the conformal gauge \ci{rtt}.}
Namely, we will set
\be
\int \frac{d^2p}{(2\pi)^2} (p^2)^n = 0\,, \ \ \  \qquad n \ge 0\,.
\ee
Introducing the notations \ci{rt1,rtt}
\be
{\rm I }[m^2]&=&\int \frac{d^2p}{(2\pi)^2}\frac{1}{(p^2+m^2)}\\
{\rm I }[m_1^2,m_2^2,m_3^2]&=&\int \frac{d^2p\, d^2q \,d^2r}{(2\pi)^4}
\frac{\delta^2(p+q+r)}{(p^2+m_1^2)(q^2+m_2^2)(r^2+m_3^2)}\,,
\ee
we obtain for the above integral
\be
\ha \int d^2p\, d^2q\, d^2r\,\delta^{(2)}(p+q+r) 
\frac{(1+4q_1^2)(1+4r_1^2)}{(p^2+1)(q^2+\ha)(r^2+\ha)}=-\fo\,\rm{I}[1,\ha,\ha]\,.
\ee
This integral is proportional to the Catalan's constant in \rf{two}
since in general 
\be \la{kow}
{\rm I }[2m^2,m^2,m^2]=\frac{\rK}{8\pi^2 m^2} \ .  \ee
The computation of the remaining contributions is analogous. 

The second term 
in (\ref{bossunset}) gives a result proportional to $\rm{I}[1]^2$, while 
the last term turns out to vanish. When everything is put together, we obtain 
the following simple answer for the bosonic sunset diagram
\be
\label{sunset}
\G_{2\ \rm{bos. sunset}}&=&\fo\,\rm{I}[1,\ha,\ha]+\ha\, \rm{I}[1]^2\,.
\ee
Note that for non-zero masses the integral ${\rm I }[m_1^2,m_2^2,m_3^2]$ is finite, while  ${\rm I }[m^2]$ is 
logarithmically UV divergent. When any of the masses vanishes, both types of integrals exhibit IR singularities.

Let us now consider the bosonic double-bubble diagram. This is given by
\be
\label{bubble}
\G_{2 \ \rm{bos. bubble}}= \langle S^{(4)}\rangle \ , 
\ee
where $S^{(4)}$ includes the following quartic vertices,
\be
&&S^{(4)}_{\tilde\phi^2 \tilde x\tilde x^* }=4 \int dt ds\,\tilde\phi^2\, |\partial_s \tilde  x-\ha \tilde x|^2\\
&&S^{(4)}_{\tilde\phi^4}=\int dt ds\,\tilde\phi^2\   [(\partial_t\tilde\phi)^2+(\partial_s\tilde\phi)^2
+\frac{1}{6}\tilde\phi^2]\\
&&S^{(4)}_{\tilde\phi^2 y^2}= \int dt ds\,\tilde\phi^2\ [(\partial_t y^a)^2+(\partial_s y^a)^2]\\
&&S^{(4)}_{y^4}= -\fo\int dt ds\ y^a y^a\, (\partial_t y^b \partial_t y^b+\partial_s y^b \partial_s y^b)\,.
\ee
It turns out that the only non-vanishing contribution comes from the $\tilde\phi^4$-interaction, and the final result is
\be
\G_{2 \ \rm{bos. bubble}}&=&-\ha\, \rm{I}[1]^2\,.
\ee
Next, let us  consider the vertices coming from the fermionic part of the Green-Schwarz action. 
 For the sunset diagram we need the following cubic interactions,
\begin{equation}\label{ferm-3vertices}
\begin{aligned}
&S^{(3)}_{\tilde\phi\tilde\eta\theta}=-2 \mathrm{i} \int dt ds\,\tilde\eta^i\,(\rho^6)_{ij}
(\partial_s\tilde\theta^j-\ha \tilde\theta^j)\tilde\phi-\rm{h.c.}\\
&S^{(3)}_{y\tilde\eta\tilde\theta}=+\mathrm{i} \int dt ds\,\tilde\eta^i(\rho^a)_{ij}
(\partial_s\tilde\theta^j-\ha\tilde\theta^j)y^a-\rm{h.c.}\\
&S^{(3)}_{\tilde x\tilde\eta\tilde\eta}=+\int dt ds\,\tilde\eta^i\,(\rho^6)_{ij}
\tilde\eta^j(\partial_s \tilde x-\ha\tilde  x)-\rm{h.c.}\\
&S^{(3)}_{y\tilde\eta\tilde\eta}=+
\mathrm{i} \int dt ds\,\tilde\eta_i\,(\rho^{a6}){}^i{}_j\tilde\eta^j\partial_t y^a\,.
\end{aligned}
\end{equation}
The fact that the bosonic propagator is diagonal 
leads to a dramatic reduction of the number of possible 
terms. The fermionic contribution to the sunset diagram is
\be\label{ferm-sunset}
\G_{2\ \rm 
{ferm.sunset} }=  -\ha\langle S^{(3)}_{\tilde\phi\tilde\eta\tilde\theta} S^{(3)}_{\tilde\phi\tilde\eta\tilde\theta}
+S^{(3)}_{\tilde x\tilde\eta\tilde\eta}S^{(3)}_{\tilde x\tilde\eta\tilde\eta}
+S^{(3)}_{y\tilde\eta\tilde\eta}S^{(3)}_{y\tilde\eta\tilde\eta}+
S^{(3)}_{y\tilde\eta\tilde\theta}S^{(3)}_{y\tilde\eta\tilde\theta}+
2S^{(3)}_{y\tilde\eta\tilde\eta}S^{(3)}_{y\tilde\eta\tilde\theta}\rangle_{\mbox{\tiny 1PI}}\,.
\ee
As an example of a typical  calculation 
 let us  detail the analysis of $\langle S^{(3)}_{\tilde\phi\tilde\eta\tilde\theta} 
S^{(3)}_{\tilde\phi\tilde\eta\tilde\theta}\rangle_{\mbox{\tiny 1PI}}$; Wick contractions  
yield the following expression
\be
-\ha\langle S^{(3)}_{\tilde\phi\tilde\eta\tilde\theta} 
S^{(3)}_{\tilde\phi\tilde\eta\tilde\theta}\rangle_{\mbox{\tiny 1PI}}
=
-\ha(2 \mathrm{i})^2 G_{\tilde\phi\tilde\phi}(r)\left[(\mathrm{i} p_1-\ha)(\mathrm{i} q_1+\ha)A-(q^2_1+\fo)B\right]
\ee
where
\begin{equation}
\begin{aligned}
&A=\,{\rm{Tr}}\left[\, \rho^\dagger_6\;  G_{\tilde\theta^{\dagger}\tilde\eta^{\dagger}}(p)\; \rho^\dagger_6\; 
G_{\tilde\theta^{\dagger}\tilde\eta^{\dagger}}(-q)
+
\; \rho^6\;  G_{\tilde\theta\tilde\eta}(p)\; \rho^6\;  G_{\tilde\theta\tilde\eta}(-q)\right]\\
&B=\,{\rm{Tr}}\left[\; \rho^6\;  G_{\tilde\eta\tilde\eta^\dagger}(p)\; \rho^\dagger_6\;  G_{\tilde\theta^\dagger\tilde\theta}(-q)
+ \; \rho^\dagger_6\;  G_{\tilde\eta^\dagger\tilde\eta}(p)\; \rho^6\;  G_{\tilde\theta\tilde\theta^\dagger}(-q)\right]\,.
\end{aligned}
\end{equation}
The fermion propagators appearing in this expression are (proportional to) the relevant entries of the
inverse of the fermionic kinetic operator (\ref{fermion-K}), and are given by
\begin{equation}
\begin{aligned}
&G_{\tilde\theta{}^i \tilde\eta{}^j}(p)=
-\frac{p_1-{\tfrac{\mathrm{i}}{2}}}{p^2+\fo}\rho^\dagger_6\,,\qquad 
G_{\tilde\theta_i \tilde\eta_j}(p)=-
\frac{p_1-{\tfrac{\mathrm{i}}{2}}}{p^2+\fo}\rho^6\,,\\
&G_{\tilde\theta^i \tilde\theta_j}(p)=G_{\tilde\eta^i \tilde\eta_j}(p)=-\frac{p_0}{p^2+\fo}{\bf 1}_4\,.
\end{aligned}
\end{equation}
After collecting all contributions in (\ref{ferm-sunset}) 
and reducing them to scalar integrals, the final  result for  the 
fermionic sunset diagram turns out to be

\be\label{fermsunset}
\G_{2\ \rm{ferm. sunset}}=- \fo\rm{I}[\ha,\fo,\fo]+2\,\rm{I}[\fo]^2 
+ 2\,\rm{I}[\fo]\rm{I}[1]-\frac{5}{2}\,\rm{I}[\fo]\rm{I}[0]\,.
\ee
Finally,  we have to include the fermionic contributions to the double-bubble topology. It is 
easy to see that the diagram with two fermion bubbles, which,  in principle,  arises due to 
the $\tilde\eta^4$ interaction, vanishes, and so do all diagrams with an 
$\tilde\eta\tilde\eta$-loop. Then the only non-trivial contributions come from the following boson-fermion 4-vertices
\be
&&S^{(4)}_{y y\tilde\eta\tilde\theta}=-\frac{\mathrm{i}}{2} \int dt ds\, y^a y^a \tilde\eta^i\,
(\rho^6)_{ij}(\partial_s\tilde\theta^j-\ha \tilde\theta^j)-\rm{h.c.}\\
&&S^{(4)}_{\tilde \phi\tilde\phi\tilde\eta\tilde\theta}=+2 \mathrm{i} \int dt ds\,\tilde\phi^2\tilde\eta^i\,
(\rho^6)_{ij}(\partial_s \tilde\theta^j-\ha \tilde\theta^j)-\rm{h.c.}
\ee
After reduction to scalar integrals
we obtain the following result
\be
 \G_{2 \ \rm{ferm. bubble}}=-2\,\rm{I}[\fo]\rm{I}[1]+ \frac{5}{2}\,\rm{I}[\fo]\rm{I}[0]\,.
\label{1PIfermi}
\ee
Thus, the bosonic and fermionic one-particle irreducible contributions, 
(\ref{sunset}), (\ref{bubble}), (\ref{fermsunset}) and (\ref{1PIfermi}), 
sum up to a divergent 2-loop correction to the partition function. 
This  UV divergence is of $\log^2$-type   and thus   should  be  canceled by 
additional non-1PI  connected diagram contributions
 to  restore 
the expected 
2-loop finiteness of the superstring theory. 
This is indeed what happens as we 
shall see  below. 

\subsection{
Additional connected graph contribution
}

\begin{figure}
\begin{center}
\includegraphics[width=60mm]{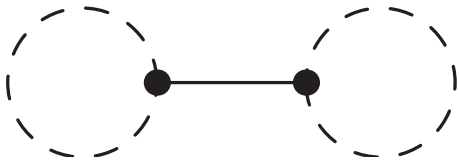}
\parbox{13cm}{\caption{The 2-loop tadpole topology. The only non-vanishing 
contribution corresponds to fermionic $\tilde\theta\tilde\eta$ bubbles, represented by 
dashed lines, connected by the  $\tilde\phi$ propagator.}
\label{tadpole}}
\end{center}
\end{figure}

So far  we have considered 
only the one-particle irreducible contributions;
however, $\ln Z_{\rm{string}}$ receives contributions 
from all connected graphs.
 In particular, at two loops we might have non-vanishing tadpole diagrams 
 of the topology given in fig. \ref{tadpole}. 
 We will see that
 such tadpole diagrams   play an important role 
for reproducing the 2-loop result  for  the cusp anomaly 
found previously in  the conformal gauge.

The (spontaneously broken) symmetries of the theory forbid single-fermion terms from appearing in the effective action.
Thus, only the 
 bosonic fields may exhibit nontrivial 1-point functions. Let us begin by discussing the contributions of bosonic loops.
Given   the structure of the bosonic 3-vertices (\ref{bose-3vertices}) and the fact that  the bosonic propagator is 
diagonal, it is easy to see that all 
2-loop graphs  with  bosonic tadpoles must 
have $\tilde\phi$ as the inner leg, which is at zero 
momentum by momentum conservation. 
Moreover, diagrams with a $\tilde\phi^2$ or a $y^2$ bubble vanish identically by 
$t \leftrightarrow s$ symmetry due to the form 
of the corresponding vertices. Thus, the only potentially non-trivial terms 
come from two $\tilde x \tilde x^{*}$ loops and the $S^{(3)}_{\tilde\phi \tilde x \tilde x^{*}}$ interaction. 
Each bubble contributes a factor
\begin{equation}
-4 \int \frac{d^2p}{(2\pi)^2} (p_1^2+\fo)G_{\tilde x \tilde x^{*}}(p)=-8 \int \frac{d^2p}{(2\pi)^2} \frac{p_1^2+\fo}{p^2+\ha}=
-4 \int \frac{d^2p}{(2\pi)^2}\ , 
\end{equation}
which is zero  in our regularization scheme, as explained above. Therefore,  we conclude that all bosonic tadpoles 
vanish.

Since all bosonic non-1PI diagrams vanish identically, the total bosonic contribution  comes 
from summing up  the expressions (\ref{sunset}) and (\ref{bubble}), i.e. is given by 
\be
 \G_{2 \ \rm{bos}} 
={1 \ov 4} \frac{2\pi}{\sqrt{\lambda}}V_2\,\rm{I}[1,\ha,\ha]\,,
\ee
where we have reinstated the explicit dependence  on the inverse of 
 the string tension $T^{-1} ={2\pi\ov \sqrt{\lambda}}$ and
 the overall two-dimensional volume factor $V_2$.

Let us  now  turn to the analysis of the fermionic tadpoles. 
Since the $\tilde\eta\tilde\eta^*$ two-point function $G_{\tilde\eta\tilde\eta^*}(p)=- {p_0\ov p^2+\fo} {\bf 1}_4$
is parity-odd, the bubble integral containing only this propagator vanishes identically. Thus, the potentially non-trivial 
tadpoles may come only  from the vertices in the first two lines of (\ref{ferm-3vertices}). 
The $S^{(3)}_{y\tilde\eta\tilde\theta}$ vertex, however, leads to a vanishing result
 since 
each bubble is proportional to ${\rm{Tr}}\left[\rho^{a} G_{\tilde \eta\tilde \theta}(p)\right] \propto \Tr\rho^{a6} = 0$. The remaining 
fermionic tadpole with a $\tilde\phi$ internal leg is, on the other hand, non-trivial and gives
\begin{equation}
-\ha\langle S^{(3)}_{\tilde\phi\tilde\eta\tilde\theta} 
S^{(3)}_{\tilde\phi\tilde\eta\tilde\theta}\rangle_{\mbox{\tiny non-1PI}}=
-\ha (2 \mathrm{i})^2 G_{\tilde\phi \tilde \phi}(0)\Big( \int \frac{d^2p}{(2 \pi)^2} (\mathrm{i} p_1-\ha)\Tr [
-
(\rho^\dagger_6) G_{\tilde\theta^\dagger \tilde\eta^\dagger}(p)
-(\rho^6) G_{\tilde\theta \tilde\eta}(p)]\Big)^2 \ , 
\end{equation}
which after reduction to scalar integrals yields
\be
\G_{2\ \rm{ferm. tadpole}}=-2\,\rm{I}[\fo]^2\,.
\ee
This $\log^2$  divergent term  is precisely what we need to cancel a similar divergent term  
in $\G_{2 \ \rm{ferm. sunset}}$, see (\ref{fermsunset}). The presence 
of the tadpole is therefore necessary to guarantee a 
finite answer for the cusp anomaly.

Let us mention that  it is   the  ``vacuum-vacuum'' transition amplitude or the background partition function \rf{ww}  that is a physical 
observable that should be UV finite. As for the 
 effective action $\Gamma$ given by the 
sum of 1PI graphs evaluated in a non-trivial background, it 
 is, in general,  UV finite only after a 
field renormalization. The presence of the tadpole for $\tilde \phi $ means 
that here one would need such a renormalization to make $\Gamma$ finite. 
We do not need to worry about this renormalization if our interest is to compute 
the  full partition function in \rf{ww}.

\

Combining all the partial results we find the answer for $\G$ at 2-loops in light-cone gauge,
\be\label{logZ}
\G_2&=&
\G_{2\ \rm{bos. sunset}}+\G_{2\ \rm{bos. bubble}}+\G_{2 \ \rm{ferm. sunset}}+\G_{2 \ \rm{ferm. bubble}}+\G_{2\ \rm{ferm. tadpole}}\nonumber\\
&=&\frac{2\pi}{\sqrt{\lambda}}V_2\Big(\fo\,\rm{I}[1,\ha,\ha]-\fo\,\rm{I}[\ha,\fo,\fo]\Big)\nonumber\\
&=& - {1 \ov 4} \frac{2\pi}{\sqrt{\lambda}}V_2\,\rm{I}[1,\ha,\ha]=-
\frac{ \rK}{8\pi\sqrt{\lambda}}V_2\,.
\ee
The result is manifestly finite and 
reproduces  the value of $a_2$ in \rf{two}.
We observe that, as in the conformal gauge calculation of \ci{rt1,rtt},
 the net effect of the fermions is to change the sign 
 of the bosonic result for $\G_2$.

We conclude that the  AdS \lcg  result is thus 
in perfect agreement with the string theory 
computation in the conformal gauge \ci{rt1,rtt} and with 
the strong-coupling prediction of the   Bethe ansatz \ci{bkk}. 

\newpage 

\subsection*{Acknowledgments}
We are grateful to
M. Kruczenski,  T. McLoughlin, R. Metsaev and D. Volin for useful discussions.
This work was supported in part by the US National Science Foundation under
DMS-0244464 (S.G.), PHY-0608114 and PHY-0855356 (R.Ro.) and PHY-0643150 (C.V.), 
  the US Department of Energy under contracts DE-FG02-201390ER40577 (OJI) (R.Ro.)
and DE-FG02-91ER40688 (C.V.), the Fundamental Laws Initiative Fund at 
Harvard University (S.G.) and the A. P. Sloan Foundation (R.Ro.).
It was also   supported by the EPSRC (R.Ri.)  and  by the PPARC (A.T.). S.G. and R.Ri. 
would like to thank the Simons Center for Geometry and Physics for hospitality during 
the 7th Simons workshop on Physics and Mathematics.


\appendix
\subsection*{Appendix A:  \ \ \ \ \ Notation}
\refstepcounter{section}
\def\theequation{A.\arabic{equation}}
\setcounter{equation}{0}

We mostly follow the  notation  of  \ci{mt2}
but define $x^\pm$ and $x,x^*$ without $1 \ov \sqrt 2$ factors. 
Four-dimensional indices (along the AdS boundary) are $a,b=0,1,2,3$; $SO(6)$ indices are 
 $ M,N=1,...,6$; $SU(4)$ indices are $i,j=1,2,3,4$.  
 For the fermionic variables  we have 
  $ \theta_i^\dagger = \theta^i\,,$ $
\eta_i^\dagger  = \eta^i\ , $
$
\theta^2 \equiv \theta^i\theta_i\,, \ 
\eta^2 \equiv \eta^i\eta_i\,.$

The matrices $\rho^M$ are off-diagonal blocks of the six-dimensional Dirac matrices
in chiral representation:
\bea
&&\rho_{ij}^M =- \rho_{ji}^M\,, \ \ \ \ \ 
   (\rho^M)^{il}\rho_{lj}^N + (\rho^N)^{il}\rho_{lj}^M
 =2\delta^{MN}\delta_j^i\,, \ \ \ \ \
  (\rho^M)^{ij}\equiv  -
 (\rho_{ij}^{M})^* \,
 \\
  &&
  \rho^{MN}{}^i{}_{ j} =\frac{1}{2}[  (\rho^M)^{il}\rho_{lj}^N
- (\rho^N)^{il}\rho_{lj}^M ]    \ . \eea
One  can
choose  the following explicit  representation for the
$\rho^M_{ij} $ matrices
\be
\rho^1_{ij}&=&\left(\begin{matrix}0&1&0&0\\-1&0&0&0\\0&0&0&1\\0&0&-1&0
\end{matrix}\right)\,,\qquad
\rho^2_{ij}=\left(\begin{matrix}0&\mathrm{i}&0&0\\-\mathrm{i}&0&0&0\\0&0&0&-\mathrm{i}\\0&0&\mathrm{i}&0
\end{matrix}\right)\,,\qquad
\rho^3_{ij}=\left(\begin{matrix}0&0&0&1\\0&0&1&0\\0&-1&0&0\\-1&0&0&0
\end{matrix}\right)\,,\nonumber \\
\rho^4_{ij}&=&\left(\begin{matrix}0&0&0&-\mathrm{i}\\0&0&\mathrm{i}&0\\0&-\mathrm{i}&0&0\\ \mathrm{i}&0&0&0
\end{matrix}\right)\,,\qquad
\rho^5_{ij}=\left(\begin{matrix}0&0&\mathrm{i}&0\\0&0&0&\mathrm{i}\\-\mathrm{i}&0&0&0\\0&-\mathrm{i}&0&0
\end{matrix}\right)\,,\qquad
\rho^6_{ij}=\left(\begin{matrix}0&0&1&0\\0&0&0&-1\\-1&0&0&0\\0&1&0&0
\end{matrix}\right)\,,\nonumber
\ee
As usual, their explicit form is not needed to carry out the calculations described in the text.
We found it convenient however to use at times the representation described here.

\newpage

\small


\end{document}